# Direct Positioning with Channel Database Assistance

Laurence Mailaender
Radio Algorithm Research
Huawei Technologies
Bridgewater, NJ USA
laurence.mailaender@huawei.com

Arkady Molev-Shteiman
Radio Algorithm Research
Huawei Technologies
Bridgewater, NJ USA
arkady.molev.shteiman@huawei.com

Xiao-Feng Qi
Radio Algorithm Research
Huawei Technologies
Bridgewater NJ USA
xiao.feng.qi@huawei.com

*Abstract*—When we have knowledge of the positions of nearby walls and buildings, estimating the source location becomes a very efficient way of characterizing and estimating a radio channel. We consider localization performance with and without this knowledge. We treat the multipath channel as a set of 'virtual receivers' whose positions can be pre-stored in a channel database. Using wall knowledge, we develop a generalized MUSIC algorithm that treats the wall reflection parameter as a nuisance variable. We compare this to a classic MVDR direct positioning algorithm that lacks wall knowledge. In a simple scenario, we find that lack of wall knowledge can increase location error by 7-100x, depending on the number of antennas, SNR, and true reflection parameter. Interestingly, as the number of antennas increases, the value of wall knowledge decreases.

## I. INTRODUCTION

As the number of antennas increases in modern Massive MIMO communications, channel estimation becomes increasingly burdensome. In fact, channel estimation may dominate the receiver algorithmic complexity, impose large signaling overhead, and become the prime limiting factor in the overall system capacity [1].

Estimation error generally increases when more uncorrelated parameters have to be estimated from the same data record. In MIMO systems it is commonly assumed that the channel between an $N$ element array and a single antenna user is characterized by $N*L$ independent parameters (where $L$ is the number of multipaths). Estimation error is therefore a serious performance issue for Massive MIMO where $N$ may be 100, 1000, or more. A step forward is achieved if the channel can instead be modeled as a sum of $B$ directional beams. In such cases the number of parameters to be estimated is dramatically reduced to $4*B$ (assuming an angle, delay, and complex amplitude for each beam). Such beam-domain techniques have been shown to offer powerful performance gains [2] [3]. Taking this concept even further, it is possible to claim that if the radio environment were known exactly, the channel becomes completely deterministic given the user and array locations, meaning there are in fact only 3 parameters to estimate (the user's $x,y,z$ position) regardless of the number of radio antennas! In other words, the user's position alone will determine the number of multipaths, their angles of arrival, and the relative phases and delays, as in a ray-tracing model. A more realistic approach is to model the channel based on user position and one complex amplitude per reflected path. This concept holds the potential for great improvements in capacity, if the user location and the radio environment (walls, buildings, etc.) can be known to sufficient precision. We anticipate that this knowledge of local scatterers can be built into a 'database' which contains information learned from the local environment.

The above discussion motivates an investigation of techniques for estimating and exploiting the user's location to improve channel estimation (see also our companion paper [4]). Traditionally, user location is established by extracting and processing certain features of the signal received at a group of sensors [5], such as the Time-of-Arrival, Time-Difference-of-Arrival, Angle-of-Arrival, Received Signal Strength, etc. More recently, the Direct Positioning technique has emerged as a way to determine the user's position from the interference pattern at the various sensors.

In [6], Direct Positioning (DP) of a single source is accomplished by computing a MUSIC spectrum at discrete points in a search grid using Virtual Sources to account for multipath reflections. Multiple arrays are processed with full or partial coherence. This is expanded to a wideband approach in [7]. A Maximum Likelihood approach in Line-of-Sight (LOS) channels is taken in [8], which accounts for multiple users but not multipath. The MVDR algorithm is used to locate multiple users in LOS channels [9], which pointed out the MVDR has an advantage over ML in not needing to know the number of users. A wideband MUSIC-based technique is also used in LOS channels [10] and an algorithm for non-coherently combining the APs is given. The user and virtual transmitter positions are estimated simultaneously in [11]. Databases and location are used to support higher-layer aspects in [12].

Our paper considers DP for both MUSIC and MVDR in a fully-coherent, multi-user, multipath case. The main contributions of this paper are:

*1)* Multipath channels are efficiently characterized by a new 'virtual receiver' description of the channel which can be incorporated into an online database. This has numerous advantages, explained below.

*2)* A new localization algorithm based on generalized MUSIC is presented, which treats the unknown amplitude of the multipath signals as nuissance variables, and,

*3)* Simulations of localization performance with and without the knowledge of the local reflector positions (the 'database') let us characterize the value of this knowledge, as a function of the number of antennas and strength of the multipath.

## II. MULTISINK AND DATABASE

### A. Virtual Transmitters or Receivers?

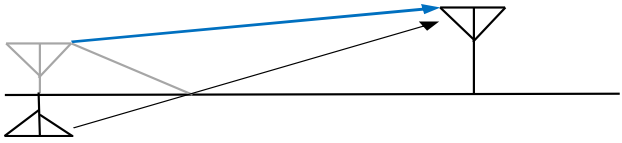

**Figure 1: Ground Bounce as Virtual Source**

The concept of modeling multipath as originating from an additional 'virtual source' is well known in the engineering literature, for example, to model the ground bounce (Figure 1) in mobile radio channels [13]. This single bounce idea is easily extended to more complex propagation scenarios where each wall, building, etc. requires an additional virtual source to model its contribution to the received signal.

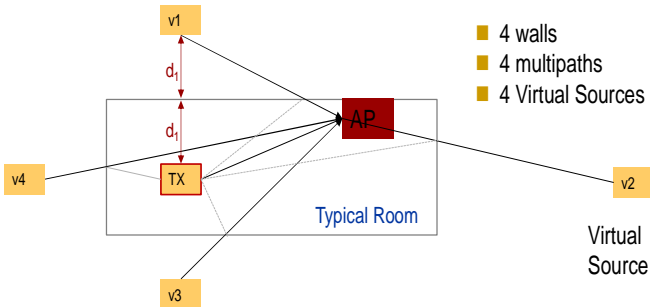

**Figure 2: Virtual Sources**

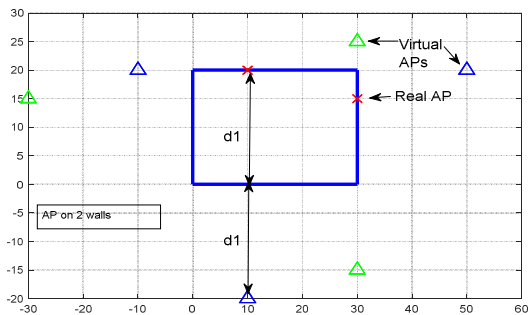

**Figure 3: Virtual Sinks**

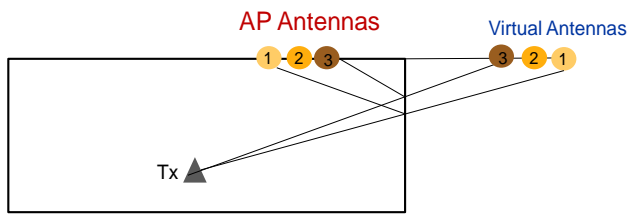

**Figure 4: Virtual Array Antennas**

In Figure 2, we see a simple rectangular room with a single access point (AP) on one wall. The transmitted signal (Tx) arrives on a direct path, and four reflected paths from the walls.

Each signal arrives *as if* from a virtual source located outside the room, at a distance that corresponds to the Tx distance to that same wall. Multiple bounces can also be handled in a like manner by adding more sources. In other words, multipath signals can be modeled as 'multi-source' signals.

This same principle can be used to model the multipath channel as a 'multisink' signal, or a set of 'virtual receivers' (sinks) as seen in Figure 3. Here a new virtual sink is added for each wall generating significant multipath. Seen are two APs (red X) on two walls. Each generates three mirror (or virtual) sinks, reflected outside the room (the blue and green triangles represent mirror sinks from the first and second AP, respectively). The mirror sinks are again located outside the room at a distance ($d_1$) equal to the original distance to the wall.

Note that the order of the antennas on mirror APs is reversed, which compensates for the apparent change in angle between the original and virtual arrays (Figure 4). The sum of all the paths at the virtual arrays models what arrives at the single real array.

### B. Benefits of Virtual Receiver Approach

Previous authors have used the Virtual Source approach to model multipath, we believe we are the first to extend this to Virtual Receivers (Sinks). This brings several benefits, including:

*1)* All Virtual Receiver locations can be pre-computed and stored in a database. This differs from Virtual Tx, where for each mobile Tx position, new virtual source locations have to be generated. Here, once the AP is installed at a known position relative to the relevant reflectors, each reflector is accounted for by a new, fixed entry in the database.

*2)* Since the user's true position explicitly appears in the model, it is easier to generate Cramer-Rao bounds from the Virtual Receiver model, which require taking derivatives with respect to the user position.

### C. Database Concept

As mentioned above, the AP and virtual AP locations are stored in a database (for further details, see [4]). We envision the database supporting a variety of localization and communications algorithms that take advantage of *site-specific* information to increase capacity. User terminal localization is a key step to enable this capability.

The Database contains two separate sections (*Figure 5*). The Location Section contains a list of antenna positions for the AP and all Virtual APs. This information encodes the local wall positions and is independent of UE location. Wall locations may be measured from visual data, from laser range-finders, or more interestingly, can be computed on-line from the communications waveform itself by various algorithms, perhaps using Artificial Intelligence. If measurements later discover that a more distant wall is making significant contributions, a new virtual sink is simply added to the first section of the database. Note that wall positions themselves can be infered from estimating the locations of the virtual sources.

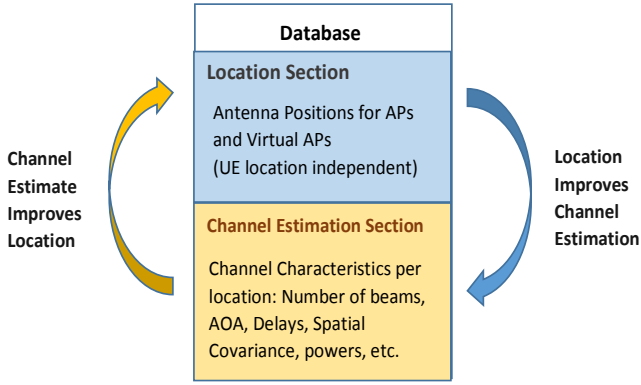

**Figure 5: Database Supporting Improved Location and Communication**

The second section characterizes the radio channel at each location. This data may be built up over time from actual channel measurements along with the positions where they were taken. Information may be statistical or deterministic. For example, we could store at a list of locations the channel impulse response generated at that location for each antenna. Or we could take a group of neighboring locations and find an expected spatial or space-time covariance for this group. A prefered approach is to store the position and amplitudes of all virtual sources associated with each location. Use of Virtual Sinks enables a moving mmW terminal to be informed about future look directions associated with individual APs.

This database will enable powerful new communications techniques that learn the local environment. As explained in the Introduction, knowledge of user locations can partially or completely determine the radio channel characteristics. In a first step, we use the database to determine the user position. The position is translated into a channel estimate using the second part of the database, yielding lower channel estimation error and higher capacity. More accurate channel estimates can in turn be used as priors to improve the location estimate, in a virtuous cycle.

Next we investigate the performance of specific location algorithms that utilize the first part of the database.

## III. DIRECT POSITONING ALGORITHMS

Direct Positioning (DP) requires a numerical search over a discrete set of possible user positions in an area, which can be computationally intensive. Lowering the complexity of the search through a multi-stage coarse/fine grid search appears feasible, and the search area can be centered on positions estimated by conventional techniques like TDoA. However, at present we are interested in determining the value of the radio database, rather than complexity reduction, which we leave for future work.

Assume that at the $p$-th Access Point (AP), the $N$ antennas and RF chains are calibrated and the array response is,

$$\mathbf{a}_{p,l}(x,y) = \exp\left(j\frac{2\pi}{\lambda}\sqrt{(\mathbf{x}_{p,l}-x)^2 + (\mathbf{y}_{p,l}-y)^2}\right) \quad (3.1)$$

where $\mathbf{x}_{p,l}, \mathbf{y}_{p,l}$ are vectors giving the $(x,y)$ coordinates of $N$ antennas associated with the $l$-th virtual receiver for the $p$-th AP. This 'near field' array response can be trivially extended to three dimensions, and is preferable to the angle domain 'far field' description when modeling indoor positioning, as the user may be quite close to the receiving array. Note that the database of virtual receiver locations $\mathbf{x}_{p,l}, \mathbf{y}_{p,l}$ over all $l$ encodes the reflector position information while leaving the user location $(x,y)$ explicit.

The signal received at the $p$-th AP from the $k$-th user is,

$$\mathbf{r}_{p,k}(t) = \mathbf{a}_{p,1}(x,y)s_k(t) + \sum_{l=2}^{L}\Gamma_{p,l}\mathbf{a}_{p,l}(x,y)s_k(t) + \mathbf{n}_{p,k} \quad (3.2)$$

which assumes that a generic signal, $s_k(t)$, from the $k^{th}$ user arrives on a direct path to the AP and at $L$-1 virtual APs corresponding to the multipath signals. Distance-dependent pathloss has been ignored. Note that the reflection coefficient, $\Gamma$, actually depends on the wall material (permittivity, conductivity), the RF frequency, and the incident angle and polarization [13]. Hence in general, different $\Gamma$ values are needed for each virtual receiver and will be user position dependent.

Recent propagation tests through wallboard at mmW frequencies showed losses of 1-3 dB [14]. This corresponds to reflection coefficients of -3 to -7 dB. While this helps us model reality, a practical positioning algorithm should treat $\Gamma$ as an unknown nuisance variable.

The model (3.2) can be re-written in matrix form as,

$$\mathbf{r}_{p,k} = \mathbf{A}_{p,k}\mathbf{b}s_k + \mathbf{n}_{p,k}$$
$$\mathbf{A}_{p,k} = [\mathbf{a}_{p,1}, \mathbf{a}_{p,2}, \cdots \mathbf{a}_{p,L}] \quad (3.3)$$
$$\mathbf{b} = [1, \Gamma, \cdots, \Gamma]^T$$

where we dropped the $(x,y)$ notation and simplify to the case where all $\Gamma$ are assumed equal. Expanding the model to include all APs and all users,

$$\mathbf{r} = \mathbf{ABs} + \mathbf{n}$$
$$= \begin{bmatrix} \mathbf{A}_{1,1} & \cdots & \mathbf{A}_{1,k} \\ \vdots & & \vdots \\ \mathbf{A}_{p,1} & \cdots & \mathbf{A}_{p,k} \end{bmatrix} \begin{bmatrix} \mathbf{b} & 0 & 0 \\ 0 & \ddots & 0 \\ 0 & 0 & \mathbf{b} \end{bmatrix} \begin{bmatrix} s_1 \\ \vdots \\ s_k \end{bmatrix} + \begin{bmatrix} \mathbf{n}_{1,1} \\ \vdots \\ \mathbf{n}_{p,k} \end{bmatrix} \quad (3.4)$$

We will process the received vector according to a Generalized MUSIC algorithm that uses the Location part of the database. If we have knowledge of $\Gamma$, then for each hypothetical $x, y$ we can define an effective steering vector,

$$\bar{\mathbf{a}} \triangleq \begin{bmatrix} \mathbf{a}_{1,0} \\ \vdots \\ \mathbf{a}_{1,P} \end{bmatrix} + \Gamma \sum_{l=2}^{L} \begin{bmatrix} \mathbf{a}_{1,l} \\ \vdots \\ \mathbf{a}_{P,l} \end{bmatrix} \triangleq \mathbf{a}_0 + \Gamma \mathbf{a}_1 \quad (3.5)$$

and the usual MUSIC spectrum is found as,

$$S(x, y, \Gamma) = \frac{\bar{\mathbf{a}}^H \bar{\mathbf{a}}}{\bar{\mathbf{a}}^H \mathbf{W} \bar{\mathbf{a}}} \quad (3.6)$$

where $\mathbf{W} = \mathbf{V}_N^H \mathbf{V}_N$ and $\mathbf{V}_N$ are the eigenvectors of the spatial covariance associated with the noise subspace.

To avoid the three dimensional search implied by (3.6) we now extend this algorithm to the case where $\Gamma$ is an unknown nuisance variable, as follows. Define a new approximate spectrum as,

$$S'(x, y) = \max_{\Gamma} S(x, y, \Gamma) \simeq \max_{\Gamma} \frac{1}{\bar{\mathbf{a}}^H \mathbf{W} \bar{\mathbf{a}}} \quad (3.7)$$

We then choose $\Gamma$ to minimize the denominator, or,

$$\hat{\Gamma} = \arg\min_{\Gamma} (\mathbf{a}_0 + \Gamma \mathbf{a}_1)^H \mathbf{W} (\mathbf{a}_0 + \Gamma \mathbf{a}_1)$$
$$= \frac{\text{Re}\{\mathbf{a}_0^H \mathbf{W} \mathbf{a}_1\}}{\mathbf{a}_0^H \mathbf{W} \mathbf{a}_0} \quad (3.8)$$

The new MUSIC spectrum is then $S(x, y, \hat{\Gamma})$ requiring only a two-dimensional search and a computation of a new $\hat{\Gamma}(x, y)$ at each point in the spectrum.

Next, for comparison we define a location algorithm which does not use any knowledge of the wall locations. As the unmodeled multipath will disturb the estimate, the signal processing must use the array's degrees of freedom to cancel the multipath. To accomplish this we use the MVDR algorithm that was also used in [9].

The steering vector in the desired look direction is $\mathbf{a}_0$ that was defined in (3.5). We use the estimated covariance,

$$\hat{\mathbf{R}} = \frac{1}{F} \sum_{f=1}^{F} \mathbf{r}(f) \mathbf{r}^H(f) \quad (3.9)$$

with $\mathbf{r}$ an instantiation of (3.4), and the well-known MVDR weight vector is,

$$\mathbf{w}_{MVDR} = \frac{\hat{\mathbf{R}}^{-1} \mathbf{a}_0}{\mathbf{a}_0^H \hat{\mathbf{R}}^{-1} \mathbf{a}_0} \quad (3.10)$$

yielding the MVDR spectrum,

$$S_{MVDR}(x, y) = \mathbf{w}_{MVDR}^H \hat{\mathbf{R}} \mathbf{w}_{MVDR} = \frac{1}{\mathbf{w}_{MVDR}^H \hat{\mathbf{R}}^{-1} \mathbf{w}_{MVDR}} \quad (3.11)$$

We also consider a Matched Filter spectrum,

$$S_{MF}(x, y) = \mathbf{a}_0^H \hat{\mathbf{R}} \mathbf{a}_0 \quad (3.12)$$

IV. NUMERICAL RESULTS

Our numerical results focus on a basic rectangular room of size 20x30m. To simplify the results, we search over a regular grid with spacing of 0.1m and further assume that the true user position always lies somewhere on this grid. In our spectrum plots, an 'o' marks the true location of the mobiles, and a '+' shows the estimated position, taken for now to be the $K$ largest values of the spectrum (this is sub-optimal).

The first group of results illustrates basic trends, holding two users at fixed locations (7,12) and (9,13). We first consider a case with 2 APs on adjacent walls, 6 antennas per AP, ideal covariance matrices, and SNR of 20 dB. When the wall reflections are weak ($\Gamma = -30 dB$) both the MUSIC and MVDR spectrums have peaks at the correct user positions (not shown). When the multipath strength is increased ($\Gamma = -7 dB$) MUSIC finds the correct positions, while MVDR made errors (Figure 6, Figure 7). MVDR is also sensitive to the estimation error in the covariance. With N=32 antennas and ideal covariance, MVDR estimated positions correctly, but when using the estimated covariance calculated over 128 vectors, MVDR had errors (Figure 8).

The next group of results show the RMS position error averaged over uniformly chosen user locations, versus the reflection coefficient along the x-axis. Again, we have 2 users, 2 AP, and we vary the number of antennas per AP. The covariance matrix is estimated over 128 vectors. Figure 9 (N=6) shows that all the techniques give similar performance when reflections are weak. As reflections increase, the performance of MUSIC *improves* as the multipaths are useful signal when the wall database is used. Note that MUSIC performance with estimated $\Gamma$ is just as good as when known. However, MVDR and MF deteriorate noticeably with stronger reflections. We now see the value of the wall location knowledge:  At Γ=-15 dB error increases 7x without it; at Γ=-7 dB, error increases 100x without it!

Increasing to N=10 (Figure 10) shows the gap between MUSIC and MVDR/MF begins to decrease. With enough spatial degrees of freedom, even MF is able to benefit from the near orthogonalization of the multipath. Lack of wall location now increases the estimation error by 35x. So wall location info is worth a bit less as the number of antennas increases. Finally, it is observed in Figure 11 that MF may even outperform MVDR, as MF does not suffer from the estimated covariance. With a sufficient number of antennas the matched filtering gain alone is enough and inverting the estimated covariance can be harmful.

## V. CONCLUSIONS

Localization is investigated as a precursor to improved channel estimation. (See also [4].) We studied a form of MUSIC that utilizes wall location knowledge, and MVDR that lacks it and tries to suppress the unknown multipath. Comparing these tells us something about the value of the wall location knowledge for localization. We find that lack of wall location knowledge increases position estimation error by 7-100x for common wall reflectivity at mmW. In general, as the strength of multipath increases, MUSIC localization improves, while that for MVDR and MF deteriorate. However, with a sufficient number of APs and antennas per AP, MVDR and MF performance could be acceptable. We find that MF can outperform MVDR due to errors in estimating the covariance matrix. To summarize, the value of wall location is very large (more than 7x) for a reasonable number of antennas, but decreases as the number of antennas grows.

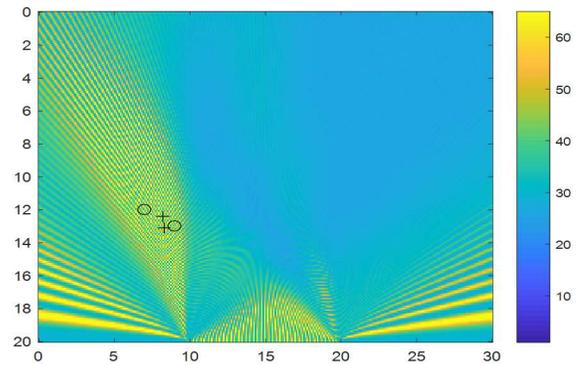

**Figure 6: MVDR Spectrum Γ=-7 dB**

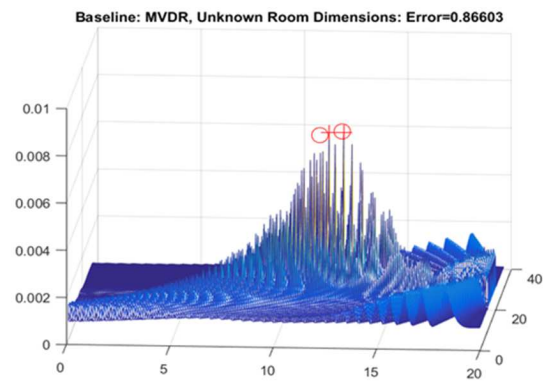

**Figure 7: Side View of MVDR Spectrum**

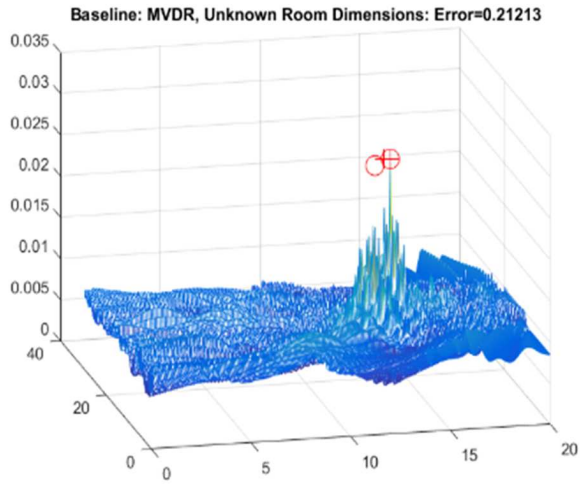

Figure 8: MVDR Estimated Covariance

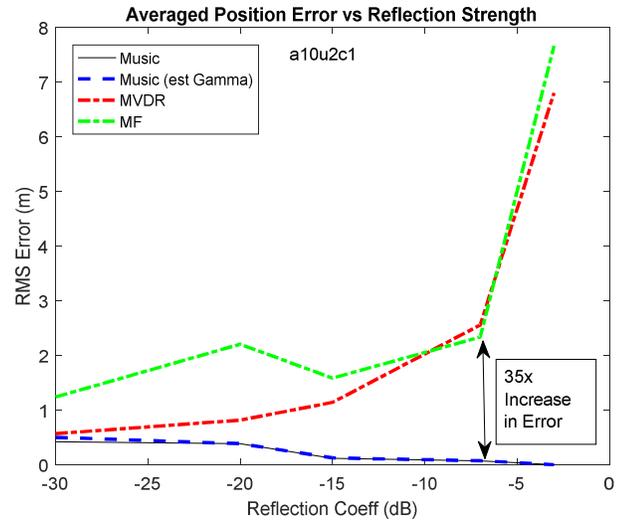

Figure 10: RMS Location Error (10 Antennas)

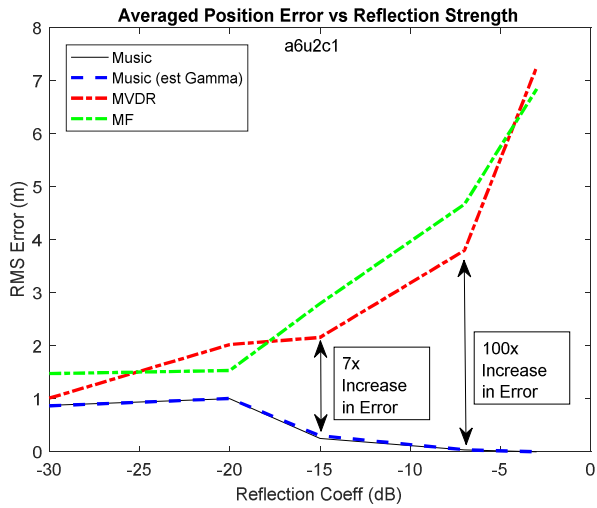

Figure 9: RMS Location Error (6 Antennas)

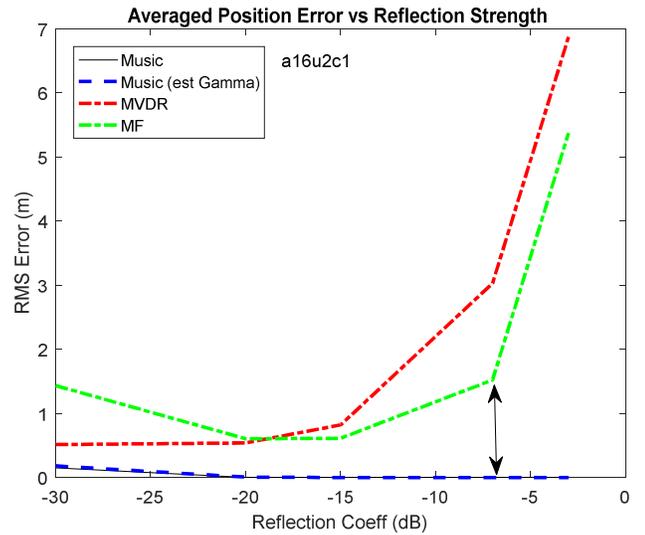

Figure 11: RMS Location Error (16 Antennas)